\documentclass[twocolumn,showpacs]{revtex4-1}
\usepackage{amsfonts,amssymb,amsmath}
\usepackage{array,dcolumn}
\usepackage{graphicx}

\begin{document}

\title{%
Emergence of second coherent regions for breathing chimera states
}

\author{Yusuke Suda${}^{1,2}$}
\author{Koji Okuda${}^{1}$}
\affiliation{%
${}^{1}$Division of Physics, Hokkaido University, Sapporo 060-0810,
Japan\\
${}^{2}$Institute for the Advancement of Higher Education, Hokkaido
University, Sapporo 060-0817, Japan
}

\date{\today}

\begin{abstract}
 Chimera states in one-dimensional nonlocally coupled phase oscillators
 are mostly assumed to be stationary, but breathing chimeras can
 occasionally appear, branching from the stationary chimeras via Hopf
 bifurcation.
 In this paper, we demonstrate two types of breathing chimeras:
 The type~I breathing chimera looks the same as the stationary chimera
 at a glance, while the type~II consists of multiple coherent regions
 with different average frequencies.
 Moreover, it is shown that the type~I changes to the type~II by
 increasing the breathing amplitude.
 Furthermore, we develop a self-consistent analysis of the local order
 parameter, which can be applied to breathing chimeras, and numerically
 demonstrate this analysis in the present system.
\end{abstract}


\maketitle

\section{Introduction}
\label{sec:Intro}

The collective dynamics of coupled nonlinear oscillators is beneficial
for understanding a wide variety of scientific phenomena
\cite{Springer.1984, Cambridge.2003}.
Chimera states can result from a symmetry breaking in a large group of
identical oscillators and have spatiotemporal patterns characterized by
the coexistence of synchronized and desynchronized oscillators.
Such a pattern was first discovered by Kuramoto and Battogtokh
\cite{NPCS.5.380} in the one-dimensional array of nonlocally coupled
complex Ginzburg-Landau (CGL) equations, which describe interacting
biological cells \cite{ProgTheorPhys.94.321}, and they introduced the
self-consistent analysis of the local mean field by phase reduction.
Then, the emergence of chimera states in the phase oscillators is
characterized by two bifurcation parameters: the phase lag parameter,
which is derived from the original parameters of the CGL equation, and
the coupling range, which is given by the diffusion factor of substance
mediating cellular interaction.
Chimera states have actively been studied and have been found in various
systems beyond the one-dimensional oscillator systems above
\cite{PhysRevLett.93.174102, IJBC.16.21, PhysRevLett.100.144102,
PhysicaD.238.1569, PhysRevE.81.065201, PhysRevE.84.015201,
Chaos.21.013112, Nonlinearity.26.2469, PhysRevE.90.022919,
IJBC.24.1440014, Nonlinearity.28.R67, JPhysA.50.08LT01, JetpLett.106.393,
Chaos.28.045101, PhysRevE.97.042212, Nonlinearity.31.R121}, with different
coupling topologies \cite{PhysRevE.69.036213, PhysRevLett.101.084103,
PhysRevLett.101.264103, PhysRevLett.104.044101, PhysRevE.93.012218},
different interaction functions \cite{Chaos.25.013106,
PhysRevE.92.060901, Nonlinearity.29.1468}, and different constituent
oscillators \cite{PhysRevLett.106.234102, PhysRevE.85.026212,
PhysRevLett.110.224101, Chaos.24.013102, Chaos.25.064401,
Chaos.25.083104, PLoSONE.12.e0187067}.
The emergence of chimera states has also been reported experimentally
\cite{NatPhys.8.658, NatPhys.8.662, PNAS.110.10563, PhysRevE.90.030902}.

When Kuramoto and Battogtokh \cite{NPCS.5.380} introduced the
self-consistent analysis of the local mean field for chimera states,
they assumed that the local mean field is time independent on the
rotating frame of the whole oscillation.
This means that the chimera state is collectively stationary.
This assumption has been used in most studies of chimeras in the
one-dimensional phase oscillator system and forms the basis of the
analytical theory \cite{NPCS.5.380, PhysRevLett.93.174102, IJBC.16.21,
PhysRevLett.100.144102, Nonlinearity.26.2469, PhysRevE.90.022919,
JPhysA.50.08LT01, JetpLett.106.393, Chaos.28.045101, PhysRevE.97.042212,
Nonlinearity.31.R121}.

A natural question arising from this assumption is whether nonstationary
chimeras exist in the one-dimensional phase oscillator system
\cite{PhysRevLett.101.084103}.
As an answer to this question, it is reported that breathing
(oscillating) chimeras can be obtained by introducing phase lag
parameter heterogeneity \cite{PhysicaD.238.1569, JetpLett.106.393,
Chaos.28.045101}.
On the other hand, we recently found that breathing chimeras can appear
even without introducing such heterogeneity \cite{PhysRevE.97.042212}.
In these previous works, it is shown that the system exhibits a Hopf
bifurcation from a stationary chimera to a breathing one.

In this paper, we study breathing chimeras in more detail.
In Sec.~\ref{sec:NumSim}, we show that two types of breathing chimeras
can be obtained by numerical simulations.
The type~I breathing chimera looks the same as the stationary chimera at
a glance, as reported in Ref.~\cite{PhysRevE.97.042212}, while the
type~II has multiple coherent regions with different average frequencies.
In Sec.~\ref{sec:Theory}, we analyze these breathing chimeras by
deriving a self-consistency equation extended for breathing chimeras and
introducing a complex function combining the average frequency and the
stability property.
In Sec.~\ref{sec:Relation}, we show that the breathing chimera can be
changed from type~I to type~II by increasing the breathing amplitude,
and then new coherent regions appear in the incoherent regions for the
type~I.
In Sec.~\ref{sec:SelfCons}, we numerically solve this self-consistency
equation.

\section{Numerical Simulation}
\label{sec:NumSim}

We consider the one-dimensional array of nonlocally coupled phase
oscillators in the continuum limit $N\to\infty$, where $N$ is the number
of oscillators.
The evolution equation of the system is given by
\begin{equation}
 \dot{\theta}(x,t) = \omega - \int^{\pi}_{-\pi} dy \, G(x-y)
  \sin[\theta(x,t) - \theta(y,t) + \alpha],
  \label{eq:PhaseOsc}
\end{equation}
with $2\pi$-periodic phase $\theta(x,t)\in[-\pi,\pi)$ on the space
$x\in[-\pi,\pi)$ under the periodic boundary condition.
The constant $\omega$ denotes the natural frequency.
The interaction between oscillators is described as the sine function
with the phase lag parameter $\alpha$ \cite{ProgTheorPhys.76.576}.
As the kernel $G(x)$ characterizing the nonlocal coupling, we use the
step kernel \cite{PhysRevE.81.065201, PhysRevE.84.015201,
Chaos.21.013112, Nonlinearity.26.2469, IJBC.24.1440014,
PhysRevE.92.060901, PhysRevE.97.042212, Nonlinearity.31.R121}
\begin{equation}
 G(x) =
  \begin{cases}
   1/(2\pi r) & (|x| \leq \pi r) \\
   0 & (|x| > \pi r),
  \end{cases}
  \label{eq:StepKernel}
\end{equation}
with $0<r\leq1$ where $r$ denotes the coupling range.
The coupling kernel is usually given by an even real function and can be
taken as, instead of the step kernel, the exponential kernel
\cite{NPCS.5.380, PhysRevLett.100.144102, JPhysA.50.08LT01,
JetpLett.106.393, Chaos.28.045101} or the cosine kernel
\cite{PhysRevLett.93.174102, IJBC.16.21, PhysicaD.238.1569,
PhysRevE.90.022919}.
For numerical simulations, we discretize $x$ into $x_j:=-\pi+2\pi j/N$
($j=0,\cdots,N-1$) and rewrite Eqs.~(\ref{eq:PhaseOsc}) and
(\ref{eq:StepKernel}) as
\begin{equation}
 \dot{\theta}_j(t) = \omega - \frac{1}{2R}
  \sum_{k=j-R}^{j+R} \sin[\theta_j(t) - \theta_k(t) + \alpha],
  \label{eq:DisPhaseOsc}
\end{equation}
where $\theta_j(t):=\theta(x_j,t)$, $R:=rN/2$ and the index $k$ is
modulo $N$.
For all the simulations of Eq.~(\ref{eq:DisPhaseOsc}), we set $\omega=0$
without loss of generality and use the fourth-order Runge-Kutta method
with time interval $\Delta t=0.01$.

\begin{figure}[tb]
 \centering
 \includegraphics[width=86truemm]{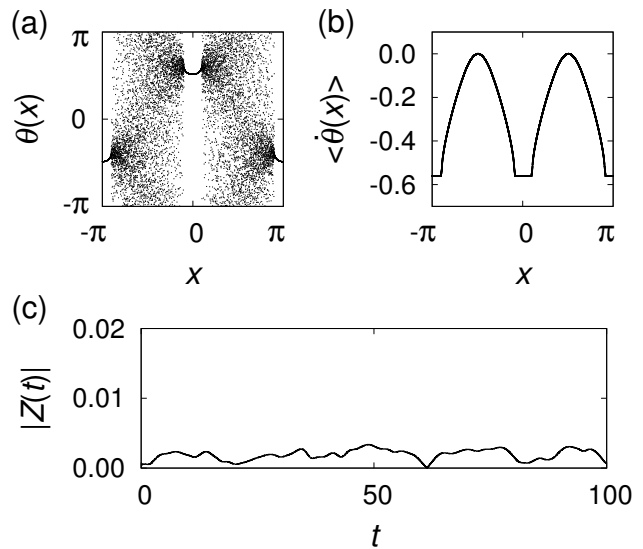}
 \caption{
 Stationary chimera state with two coherent and incoherent regions for
 Eq.~(\ref{eq:DisPhaseOsc}) with $N=100000$, $\alpha=1.480$, and
 $r=0.440$.
 (a)~The snapshot of the phase $\theta(x,t)$.
 (b)~The profile of the average frequency
 $\langle\dot{\theta}(x,t)\rangle$ with $T=2000$.
 (c)~Time evolution of the global order parameter $|Z(t)|$.
 Figures~(a) and (b) are plotted once every 10 oscillators.
 }
 \label{fig:Stationary}
\end{figure}

Chimera states for Eq.~(\ref{eq:PhaseOsc}) are characterized by the
coexistence of coherent and incoherent regions.
For example, Fig.~\ref{fig:Stationary} shows a chimera state with two
coherent and incoherent regions.
In the coherent region, the oscillators are synchronized with each other
at a constant average frequency, while the oscillators in the incoherent
region are drifting at continuously varying average frequencies.
The average frequency is numerically defined as
\begin{equation}
 \langle \dot{\theta}(x,t) \rangle
  := \frac{1}{T} \int_{0}^{T} dt' \, \dot{\theta}(x,t'),
  \label{eq:AveFreq}
\end{equation}
with the measurement time $T$ after a sufficiently long transient time.
In the following, $\langle\cdot\rangle$ denotes the time average
quantity.

\begin{figure}[tb]
 \centering
 \includegraphics[width=86truemm]{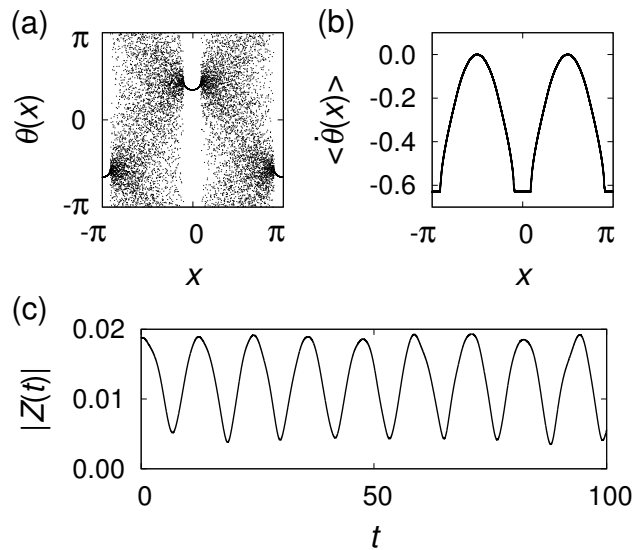}
 \caption{
 The type~I breathing chimera for Eq.~(\ref{eq:DisPhaseOsc}) with
 $N=100000$, $\alpha=1.480$, and $r=0.360$.
 All figures show the same quantities as those in
 Fig.~\ref{fig:Stationary}.
 }
 \label{fig:Type1}
\end{figure}

While the chimera state in Fig.~\ref{fig:Stationary} is a stationary
state, we have found breathing chimeras with two coherent and incoherent
regions \cite{PhysRevE.97.042212}, as shown in Fig.~\ref{fig:Type1},
which we call the type~I breathing chimera below.
Though the stationary and the type~I breathing chimeras have
very similar appearance of the phase snapshot, they can be distinguished
by observing the time evolution of the global order parameter $|Z(t)|$,
defined as
\begin{equation}
 Z(t) := \frac{1}{2\pi} \int^{\pi}_{-\pi} dy \, e^{i\theta(y,t)},
  \label{eq:GlobalOrder}
\end{equation}
which denotes the synchronization degree of all the oscillators.
For $|Z(t)|=1$, all the oscillators are completely synchronized in
phase, and otherwise for $0\leq|Z(t)|<1$.
In the present case, $|Z(t)|$ becomes nearly zero for
both of the stationary and the type~I breathing chimeras,
but the time evolutions are different.
For stationary chimeras, $|Z(t)|$ is time-independent.
Figure~\ref{fig:Stationary}(c) denotes a small fluctuation around zero
and can be regarded as nearly satisfying $|Z(t)|=0$.
For breathing chimeras with sufficiently large $N$, however, $|Z(t)|$
oscillates periodically, as shown in Fig.~\ref{fig:Type1}(c).

\begin{figure}[tb]
 \centering
 \includegraphics[width=86truemm]{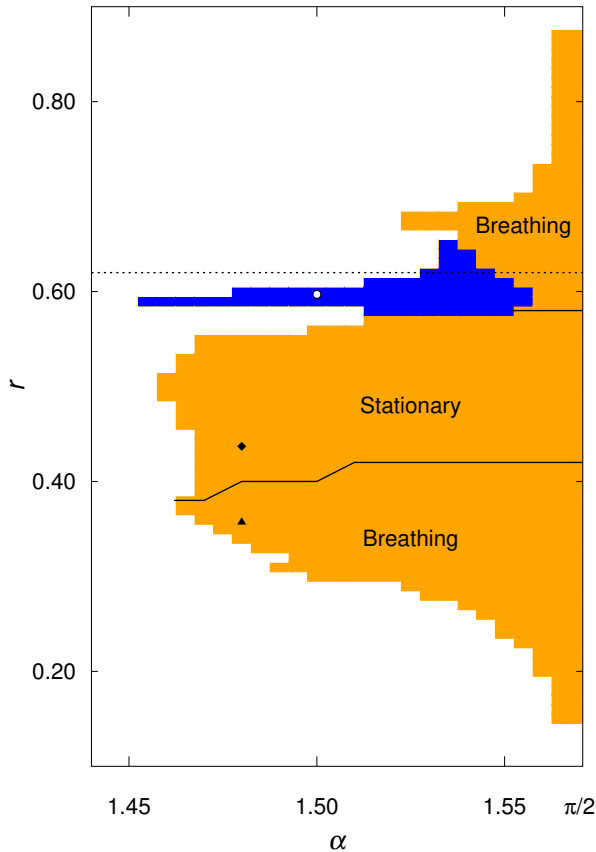}
 \caption{
 Stability region of chimera states obtained by the numerical simulation
 of Eq.~(\ref{eq:DisPhaseOsc}).
 There appear the stationary and the type~I breathing chimeras in the
 orange region and the type~II breathing chimeras in the blue region.
 Black solid lines denote the Hopf bifurcation points
 \cite{PhysRevE.97.042212}.
 The black diamond, the black triangle, and the white circle denote the
 parameter values of Figs.~\ref{fig:Stationary}, \ref{fig:Type1}, and
 \ref{fig:Type2}, respectively.
 The horizontal dotted line denotes the parameter $r=0.620$ discussed in
 Sec.~\ref{sec:Relation}.
 }
 \label{fig:StabRegion}
\end{figure}

In our simulations of Eq.~(\ref{eq:DisPhaseOsc}), the stationary and the
type~I breathing chimeras with two coherent and incoherent regions are
obtained in the orange region in Fig.~\ref{fig:StabRegion}.
In our previous work \cite{PhysRevE.97.042212}, we showed that the
breathing chimera branches from the stationary one via supercritical
Hopf bifurcation.
The bifurcation points are indicated by
black solid lines
in Fig.~\ref{fig:StabRegion}.
We previously showed only the bifurcation points at $r\simeq0.400$ by
the linear stability analysis of the stationary chimera
\cite{PhysRevE.97.042212}.
However, we have found the other bifurcation points at $r\simeq0.580$
by the same method as before.
Breathing chimeras are also found in two interacting populations of
globally coupled phase oscillators, where the global order parameter of
a desynchronized population oscillates temporally
\cite{PhysRevLett.101.084103, PhysRevE.93.012218}.

\begin{figure}[tb]
 \centering
 \includegraphics[width=86truemm]{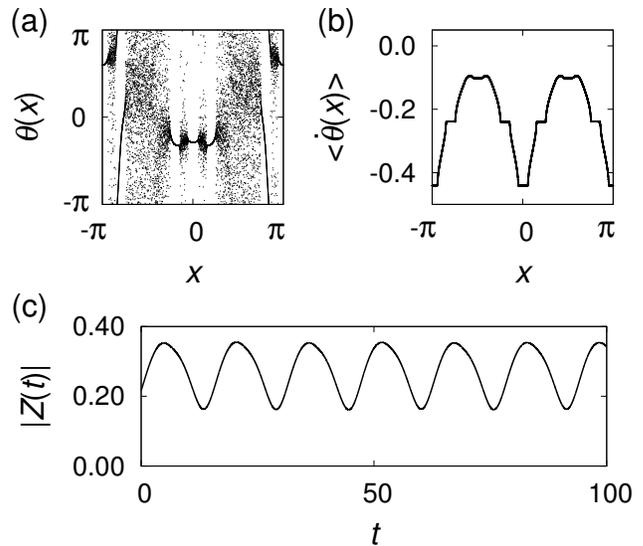}
 \caption{
 The type~II breathing chimera for Eq.~(\ref{eq:DisPhaseOsc}) with
 $N=100000$, $\alpha=1.500$, and $r=0.600$.
 All figures show the same quantities as those in
 Fig.~\ref{fig:Stationary}.
 }
 \label{fig:Type2}
\end{figure}

In addition to the type~I breathing chimera, we have numerically found the
type~II breathing chimera characterized by two kinds of coherent regions
with different average frequencies, as shown in Fig.~\ref{fig:Type2}.
The first coherent regions around $x=0$ and $x=\pm\pi$ in
Fig.~\ref{fig:Type2}(a) are similar to the coherent regions of the
stationary or the type~I breathing chimera; that is, they are always
separated from each other by the phase almost exactly $\pi$.
The second coherent regions lie near each first coherent region and have
a different average frequency from it.
Such type~II breathing chimeras with multiple coherent regions are also
observed in the system with phase lag parameter heterogeneity
\cite{JetpLett.106.393, Chaos.28.045101}.
The stability region of the type~II breathing chimeras is shown as the
blue region in Fig.~\ref{fig:StabRegion}.
In our numerical simulations, we did not find the bistable region of the
types~I and II.
In this paper, we focus on these two types of breathing chimeras and aim
to understand them theoretically.

\section{Theory of Breathing Chimeras}
\label{sec:Theory}

In this section, we study the properties common to the two types of
breathing chimeras.
First, we define the local order parameter and the local mean field.
The local order parameter \cite{PhysRevE.90.022919}
\begin{equation}
 z(x,t) := \lim_{\eta\to0+} \frac{1}{2\eta}
  \int^{x+\eta}_{x-\eta} dy \, e^{i\theta(y,t)},
  \label{eq:LocalOrder}
\end{equation}
which satisfies $0\leq|z(x,t)|\leq1$, is similar to the global order
parameter given by Eq.~(\ref{eq:GlobalOrder}) in quality, and $|z(x,t)|$
denotes the synchronization degree of oscillators in the neighborhood of
a point $x$.
In the case of chimera states, $|z(x,t)|=1$ implies that the oscillator
at $x$ belongs to a coherent region, and otherwise an incoherent region.
We assume that the number of oscillators contained in the integral of
Eq.~(\ref{eq:LocalOrder}) tends to infinity in the continuum limit
$N\to\infty$.
The local mean field \cite{NPCS.5.380} is defined as
\begin{equation}
 Y(x,t) := \int^{\pi}_{-\pi} dy \, G(x-y) \, e^{i\theta(y,t)}.
  \label{eq:MeanField}
\end{equation}
Then, Eq.~(\ref{eq:PhaseOsc}) is rewritten as
\begin{equation}
 \dot{\theta}(x,t) = \omega
  - {\rm Im}[e^{i\alpha} e^{i\theta(x,t)} Y^*(x,t)],
  \label{eq:PhaseOsc_2}
\end{equation}
where the symbol $*$ denotes the complex conjugate.
Equation~(\ref{eq:PhaseOsc_2}) suggests a physical picture in which each
independent phase oscillator is driven by the local mean field $Y(x,t)$.
Using the local order parameter $z(x,t)$, Eqs.~(\ref{eq:GlobalOrder}) and
(\ref{eq:MeanField}) are rewritten as
\begin{equation}
 Z(t) = \int^{\pi}_{-\pi} dy \, z(y,t),
  \label{eq:GO_LO}
\end{equation}
\begin{equation}
 Y(x,t) = \int^{\pi}_{-\pi} dy \, G(x-y) \, z(y,t).
  \label{eq:MF_LO}
\end{equation}
In the continuum limit, phase oscillators described as
Eq.~(\ref{eq:PhaseOsc_2}) can be regarded as interacting subpopulations
of globally coupled infinite oscillators in the neighborhood of $x$
\cite{Chaos.21.013112}.
Then, we can obtain the evolution equation of $z(x,t)$ as
\begin{equation}
 \dot{z}(x,t) = i \omega z(x,t)
  + \frac{1}{2} e^{-i\alpha} Y(x,t)
  - \frac{1}{2} e^{i\alpha} z^2(x,t) Y^*(x,t),
  \label{eq:z-Eq}
\end{equation}
by the method in Refs.~\cite{PhysRevLett.101.264103, Chaos.21.013112}
using the Watanabe-Strogatz approach \cite{PhysicaD.74.197}.
We can also define the local order parameter by using a probability
density function of phase \cite{PhysicaD.238.1569, Nonlinearity.26.2469,
Nonlinearity.31.R121}.
In that case, Eq.~(\ref{eq:z-Eq}) can be obtained from the Ott-Antonsen
ansatz \cite{Chaos.18.037113, Chaos.19.023117}.

If chimera states are stationary, the local order parameter takes the
form
\begin{equation}
 z(x,t) = z_{\rm st}(x) \, e^{i\Omega t},
  \label{eq:Stat-LO}
\end{equation}
with the frequency $\Omega$ of the rotating frame, which we may regard as
the definition of ``stationary'' for chimera states.
Then, the local mean field is also obtained as
\begin{equation}
 Y(x,t) = Y_{\rm st}(x) \, e^{i\Omega t},
  \label{eq:Stat-MF}
\end{equation}
from Eq.~(\ref{eq:MF_LO}).
Using Eqs.~(\ref{eq:Stat-LO}) and (\ref{eq:Stat-MF}),
Eq.~(\ref{eq:z-Eq}) is rewritten as
\begin{equation}
 0 = i \Delta z_{\rm st}(x)
  + \frac{1}{2} e^{-i\alpha} Y_{\rm st}(x)
  - \frac{1}{2} e^{i\alpha} z_{\rm st}^2(x) Y_{\rm st}^*(x),
 \label{eq:Stat-z-Eq}
\end{equation}
where $\Delta:=\omega-\Omega$.
When Eq.~(\ref{eq:Stat-z-Eq}) is regarded as a quadratic equation with
respect to $z_{\rm st}(x)$, the stable solution satisfying
$0\leq|z(x,t)|\leq1$ is
\begin{equation}
 z_{\rm st}(x) = e^{-i\alpha} [i\Delta - g(x)]/Y_{\rm st}^*(x),
  \label{eq:Sol_Stat-z}
\end{equation}
\begin{equation}
 g(x) :=
  \begin{cases}
   -|\Delta| \sqrt{(|Y_{\rm st}(x)|/\Delta)^2 - 1}
    & [|\Delta| \leq |Y_{\rm st}(x)|] \\
   i\Delta \sqrt{1 - (|Y_{\rm st}(x)|/\Delta)^2}
    & [|\Delta| > |Y_{\rm st}(x)|],
  \end{cases}
  \label{eq:EssSpec}
\end{equation}
where $|\Delta|\leq|Y_{\rm st}(x)|$ and $|\Delta|>|Y_{\rm st}(x)|$
correspond to coherent and incoherent regions, respectively, and in
Eq.~(\ref{eq:Stat-Stab}) we confirm that this solution in
Eq.~(\ref{eq:Sol_Stat-z}) satisfies the local stability condition.
Moreover, taking its convolution with the coupling kernel $G(x)$, we can
obtain the self-consistency equation of $Y_{\rm st}(x)$ as
\begin{equation}
 Y_{\rm st}(x) = e^{-i\alpha} \int^{\pi}_{-\pi} dy \, G(x-y)
  [i\Delta - g(y)]/Y_{\rm st}^*(y),
  \label{eq:SelfCons}
\end{equation}
which agrees with the equation derived by Kuramoto and Battogtokh
\cite{NPCS.5.380}.

For breathing chimeras, instead of Eq.~(\ref{eq:Stat-LO}), we assume
that the local order parameter takes the form
\begin{equation}
 z(x,t) = \sum_{k=-\infty}^{\infty}
  z_k(x) \, e^{i(\Omega + k\delta)t},
  \label{eq:Breath-LO}
\end{equation}
introducing the breathing frequency $\delta$ in addition to the frequency
$\Omega$ of the rotating frame.
We take the sign of $\delta$ in accordance with $\Delta$; for example,
when $\Delta>0$, we set $\delta>0$.
Then,
\begin{equation}
 Y(x,t) = \sum_{k=-\infty}^{\infty} Y_k(x) \, e^{i(\Omega+k\delta)t},
  \label{eq:Breath-MF}
\end{equation}
\begin{equation}
 Y_k(x) = \int^{\pi}_{-\pi} dy \, G(x-y) \, z_k(y),
  \label{eq:MF-k_LO-k}
\end{equation}
are also obtained from Eq.~(\ref{eq:MF_LO}).
Equation~(\ref{eq:Breath-LO}) is equivalent to the Fourier expansion of
$z(x,t)$ and includes the stationary solution where
$z_0(x)=z_{\rm st}(x)$ and $z_{k\neq0}(x)=0$.
Substituting Eqs.~(\ref{eq:Breath-LO}) and (\ref{eq:Breath-MF}) into
Eq.~(\ref{eq:z-Eq}), we obtain the following equation for each $k$:
\begin{eqnarray}
 0 &=& i \Delta_k z_k(x)
  + \frac{1}{2} e^{-i\alpha} Y_k(x) \nonumber \\
  & &- \frac{1}{2} e^{i\alpha}
  {\displaystyle \sum_{l+m-n=k}} z_l(x) z_m(x) Y_n^*(x),
  \label{eq:Breath-z-Eq}
\end{eqnarray}
where $\Delta_k:=\omega-\Omega-k\delta$.
Similarly to stationary chimeras, we also regard
Eq.~(\ref{eq:Breath-z-Eq}) as a quadratic equation with respect to
$z_k(x)$ and obtain the solution
\begin{equation}
 z_k(x) = [ B_k(x) + \{ {B_k}^2(x)-A_k(x)C_k(x) \}^{\frac{1}{2}} ]
  / A_k(x),
  \label{eq:Sol_Breath-z}
\end{equation}
\begin{equation}
 A_k(x) := e^{i\alpha} Y_k^*(x),
  \label{eq:Quad-Eq_A}
\end{equation}
\begin{equation}
 B_k(x) := i\Delta_k - e^{i\alpha} \sum_{l \neq k} z_l(x) Y_l^*(x),
  \label{eq:Quad-Eq_B}
\end{equation}
\begin{equation}
 C_k(x) := -e^{-i\alpha} Y_k(x) + e^{i\alpha}
  \sum_{\substack{l \neq k \\ m \neq k}} z_l(x) z_m(x) Y_{l+m-k}^*(x).
  \label{eq:Quad-Eq_C}
\end{equation}
As the argument of the square root in Eq.~(\ref{eq:Sol_Breath-z}),
either one should be chosen to satisfy $|z(x,t)|\leq1$ and the stability
condition of the oscillator if it belongs to a coherent region.
We can regard Eqs.~(\ref{eq:Sol_Breath-z})-(\ref{eq:Quad-Eq_C}) as the
new self-consistency equations of the set of the complex coefficient
function $\{z_k(x)\}$ for breathing chimeras, which are discussed in
Sec.~\ref{sec:SelfCons}.

\begin{figure}[tb]
 \centering
 \includegraphics[width=86truemm]{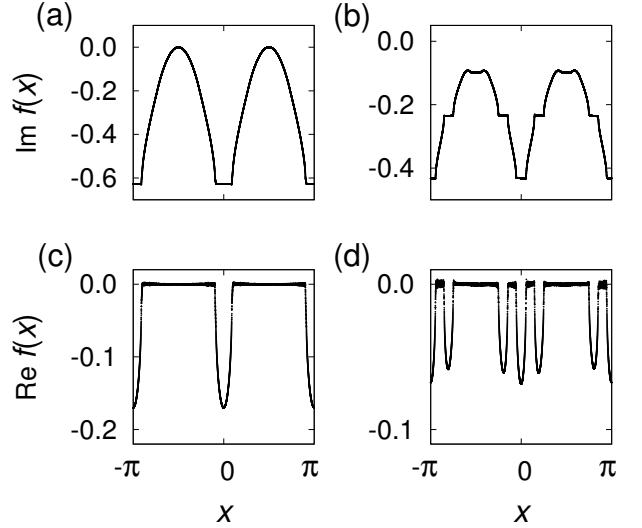}
 \caption{
 Profile of Eq.~(\ref{eq:LO-conjMF}) for the type~I [(a) and (c)] and
 the type~II [(b) and (d)] breathing chimeras corresponding to
 Figs.~\ref{fig:Type1} and \ref{fig:Type2}, respectively.
 [(a), (b)] The imaginary part denotes the average frequency.
 [(c), (d)] The real part denotes the linear stability against a small
 local perturbation, where it is negative in stable coherent regions and
 zero in neutral incoherent regions.
 All figures are depicted by computing $z_k(x)$ and $Y_k(x)$ for
 $k\in[-5,5]$ obtained by the numerical simulation of
 Eq.~(\ref{eq:DisPhaseOsc}).
 }
 \label{fig:LO-conjMF}
\end{figure}

The average frequency of breathing chimeras can be obtained by using
Eq.~(\ref{eq:Breath-LO}).
To simplify the notation, we describe the right-hand side of
Eq.~(\ref{eq:LocalOrder}) as $\mathcal{P}e^{i\theta}$ with an operator
$\mathcal{P}$ below.
$\mathcal{P}A$ means that the function $A(x)$ is averaged in the
neighborhood of a point $x$, that is,
\begin{equation}
 (\mathcal{P}A)(x) := \lim_{\eta\to0+} \frac{1}{2\eta}
  \int^{x+\eta}_{x-\eta} dy \, A(y).
  \label{eq:Operator-P}
\end{equation}
We note that the continuous functions, e.g., $Y(x)$, are not affected
by $\mathcal{P}$.
Operating $\mathcal{P}$ on Eq.~(\ref{eq:PhaseOsc_2}), we have
\begin{equation}
 (\mathcal{P} \, \dot{\theta})(x,t) = \omega
  - {\rm Im}[e^{i\alpha} z(x,t) \, Y^*(x,t) ].
  \label{eq:Cal_AveFreq_1}
\end{equation}
Note that the right-hand side of Eq.~(\ref{eq:Cal_AveFreq_1}) agrees
with the other equation obtained by the Watanabe-Strogatz approach
together with Eq.~(\ref{eq:z-Eq}) (see Eq.~(11) in
Ref.~\cite{PhysRevLett.101.264103}).
Averaging both sides of Eq.~(\ref{eq:Cal_AveFreq_1}) temporally, since
$\mathcal{P}$ and $\langle\cdot\rangle$ are commutative, we have
\begin{equation}
 \langle \dot{\theta}(x,t) \rangle = \omega
  - {\rm Im}[e^{i\alpha} \langle z(x,t) \, Y^*(x,t) \rangle].
  \label{eq:Cal_AveFreq_2}
\end{equation}
Moreover, because
\begin{equation}
 \langle z(x,t) \, Y^*(x,t) \rangle =
  \sum_{k=-\infty}^{\infty} z_k(x) \, Y_k^*(x),
  \label{eq:Cal_AveFreq_3}
\end{equation}
is established for a sufficiently long measurement time, from
Eq.~(\ref{eq:Cal_AveFreq_2}) we obtain the average frequency as the
imaginary part of the complex function
\begin{equation}
 f(x) := i\omega - e^{i\alpha}
  \sum_{k=-\infty}^{\infty} z_k(x) \, Y_k^*(x).
  \label{eq:LO-conjMF}
\end{equation}
Figures~\ref{fig:LO-conjMF}(a) and \ref{fig:LO-conjMF}(b) show the
profiles of the imaginary part of Eq.~(\ref{eq:LO-conjMF}) corresponding
to the average frequencies in Figs.~\ref{fig:Type1}(b) and
\ref{fig:Type2}(b).
All figures in Fig.~\ref{fig:LO-conjMF} are depicted by computing
$z_k(x)$ and $Y_k(x)$ for $k\in[-5,5]$ in the numerical simulation of
Eq.~(\ref{eq:DisPhaseOsc}), and then we have computed $z_k(x)$ and
$Y_k(x)$ as
\begin{equation}
 z_k(x) = \langle e^{i\theta(x,t)} e^{-i(\Omega+k\delta)t} \rangle,
  \label{eq:Num_LO-k}
\end{equation}
\begin{equation}
 Y_k(x) = \langle Y(x,t) e^{-i(\Omega+k\delta)t} \rangle,
  \label{eq:Num_MF-k}
\end{equation}
using the inverse transformation of Eqs.~(\ref{eq:Breath-LO}) and
(\ref{eq:Breath-MF}), where $\Omega$ and $\delta$ are computed
from the Fourier transform of the time series $Z(t)$ by using
Eqs.~(\ref{eq:GO_LO}) and (\ref{eq:Breath-LO}).

In addition to the average frequency, we note that the real part of
Eq.~(\ref{eq:LO-conjMF}) denotes an important property of breathing
chimeras, that is, the linear stability against a small local
perturbation.
Suppose that only the oscillator at $x$ is perturbed from $\theta(x,t)$
to $\theta(x,t)+\phi(x,t)$, where $\phi$ is small.
Then, we are allowed to regard the local mean field $Y(x,t)$ as
unchanged by that perturbation, as far as the continuum limit is
considered, since the perturbation at only one point $x$ does not
affect the integrated value $Y(x,t)$ in that limit.
From Eq.~(\ref{eq:PhaseOsc_2}), we can obtain the linear evolution
equation of $\phi(x,t)$ as
\begin{equation}
 \dot{\phi}(x,t) = [\partial_{\theta}V(\theta,x)] \, \phi(x,t),
  \label{eq:Linear-Eq_1}
\end{equation}
\begin{equation}
 \partial_{\theta} V(\theta,x)
  = -{\rm Re}[e^{i\alpha} e^{i\theta} Y^*(x,t)],
  \label{eq:Linear-Eq_2}
\end{equation}
where $V(\theta,x)$ denotes the right-hand side of
Eq.~(\ref{eq:PhaseOsc_2}).
When our breathing chimera is stable, the time-averaged
$\langle\partial_{\theta}V(\theta,x)\rangle$ should be nonpositive.
We act with the operator $\mathcal{P}$ on Eq.~(\ref{eq:Linear-Eq_2})
and average the result over time, as Eqs.~(\ref{eq:Cal_AveFreq_1}) and
(\ref{eq:Cal_AveFreq_2}).
Moreover, using Eq.~(\ref{eq:Cal_AveFreq_3}), we finally obtain
\begin{equation}
 \langle \partial_{\theta} V(\theta,x) \rangle =
  -{\rm Re}[e^{i\alpha} \sum_{k=-\infty}^{\infty} z_k(x) \, Y_k^*(x)],
  \label{eq:Linear-Eq_3}
\end{equation}
which is equivalent to the real part of Eq.~(\ref{eq:LO-conjMF}).
Here, we assumed that $\langle\partial_{\theta}V(\theta,x)\rangle$
is a continuous function with respect to $x$, namely, which is not
affected by $\mathcal{P}$.
Figures~\ref{fig:LO-conjMF}(c) and \ref{fig:LO-conjMF}(d) show the
profiles of the real part of Eq.~(\ref{eq:LO-conjMF}).
In the coherent regions, the real part of Eq.~(\ref{eq:LO-conjMF}) is
negative, while that is zero in the incoherent regions.
This implies that the oscillators are locally stable in the coherent
regions and neutral in the incoherent regions.

For stationary chimeras, Eq.~(\ref{eq:LO-conjMF}) is
\begin{equation}
 f(x) = i\omega - e^{i\alpha} z_{\rm st}(x) \, Y_{\rm st}^*(x).
  \label{eq:Stat-LOcMF}
\end{equation}
From Eqs.~(\ref{eq:Sol_Stat-z}) and (\ref{eq:EssSpec}), we obtain
$f(x)=i\Omega+g(x)$, therefore
\begin{equation}
 {\rm Im} f(x) =
  \begin{cases}
   \Omega
    & [|\Delta| \leq |Y_{\rm st}(x)|] \\
   \Omega + \Delta \sqrt{1 - (|Y_{\rm st}(x)|/\Delta)^2}
    & [|\Delta| > |Y_{\rm st}(x)|],
  \end{cases}
  \label{eq:Stat-AveFreq}
\end{equation}
which agrees with the average frequency derived by Kuramoto and
Battogtokh \cite{NPCS.5.380}.
The stability property is also obtained as
\begin{equation}
 {\rm Re} f(x) =
  \begin{cases}
   -|\Delta| \sqrt{(|Y_{\rm st}(x)|/\Delta)^2 - 1}
    & [|\Delta| \leq |Y_{\rm st}(x)|] \\
   0
    & [|\Delta| > |Y_{\rm st}(x)|].
  \end{cases}
  \label{eq:Stat-Stab}
\end{equation}
We note that the set of $g(x)$ and its complex conjugate is the
essential spectrum obtained by the linear stability analysis of the
stationary chimera \cite{Nonlinearity.26.2469, Nonlinearity.31.R121}.

\section{Relation between Two Types of Breathing Chimeras}
\label{sec:Relation}

Next, we study the relation between the two types of breathing chimeras
in this section.
In particular, we fix the parameter $r=0.620$, which corresponds to
the horizontal dotted line in Fig.~\ref{fig:StabRegion}, and compare
the two types of breathing chimeras with close parameters.
For the numerical simulation of Eq.~(\ref{eq:DisPhaseOsc}) with fixed
$r=0.620$, the emergence of the types~I and II is switched at
$\alpha\simeq1.550$; namely, the type~I is stable for
$1.550<\alpha<\pi/2$ and the type~II for $\alpha<1.550$.

\begin{figure}[tb]
 \centering
 \includegraphics[width=86truemm]{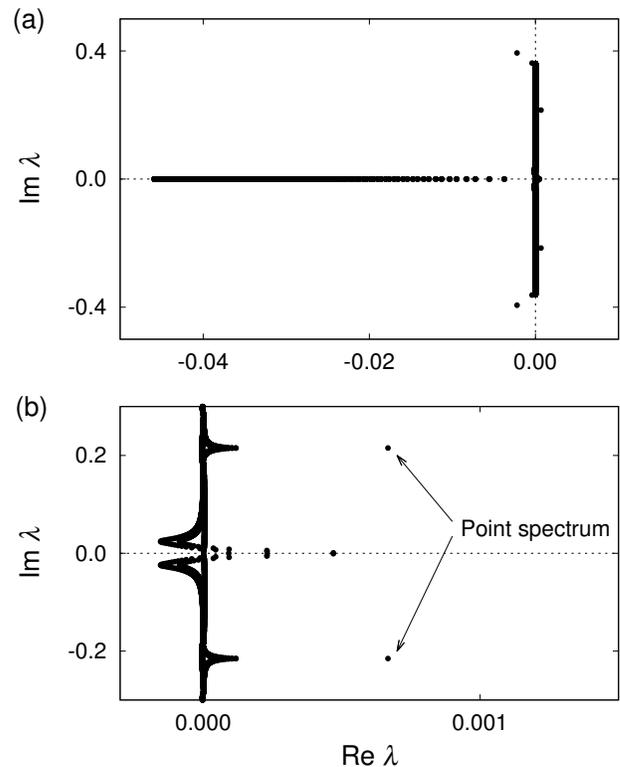}
 \caption{
 Complex eigenvalues $\lambda$ for the unstable stationary chimera state
 with $\alpha=1.549$ and $r=0.620$.
 (a)~All eigenvalues.
 (b)~The enlarged view of panel~(a).
 The dashed lines in each panel are drawn only for reference.
 }
 \label{fig:Eigenvalue}
\end{figure}

By the linear stability analysis of the stationary chimera
\cite{Nonlinearity.26.2469, PhysRevE.90.022919, JPhysA.50.08LT01,
JetpLett.106.393, Chaos.28.045101, PhysRevE.97.042212,
Nonlinearity.31.R121}, the eigenvalues characterizing the stability of
the stationary chimera can be obtained as the essential spectrum and the
point spectrum.
Then, the essential spectrum is given by the set of $g(x)$ [described as
Eq.~(\ref{eq:EssSpec})] and its complex conjugate, which consists of
pure imaginary and negative real eigenvalues, and the point spectrum
determines whether the stationary chimera is stable.
Figure~\ref{fig:Eigenvalue} shows an example of the eigenvalues $\lambda$
for an unstable stationary chimera state obtained by the numerical method
in Ref.~\cite{PhysRevE.97.042212}, where we computed the eigenvalues from
the finite (but large) sized linearized matrix obtained by discretizing
the space coordinate of Eq.~(\ref{eq:z-Eq}).
The point spectrum is a pair of the complex conjugate eigenvalues with a
positive real value and the imaginary values about $\pm0.215$.
Though there are eigenvalues with positive real values around the real
axis, they belong to the fluctuation of the essential spectrum by finite
discretization of the numerical method, and approach zero by finer
discretization \cite{PhysRevE.97.042212}.

\begin{figure}[tb]
 \centering
 \includegraphics[width=86truemm]{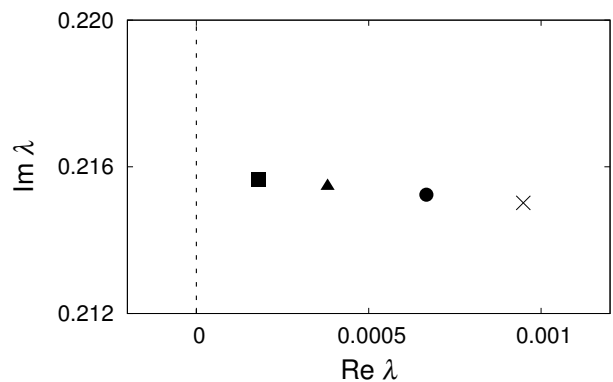}
 \caption{
 Transition of the point spectrum with a positive imaginary value for
 fixed $r=0.620$.
 The square, the triangle, the circle, and the cross denote the point
 spectrum for $\alpha=1.551$, $\alpha=1.550$, $\alpha=1.549$, and
 $\alpha=1.548$, respectively.
 The dashed line is the imaginary axis.
 }
 \label{fig:PointSpec}
\end{figure}

We numerically computed these spectra for fixed $r=0.620$, and obtained
results such that the positive real part of the point spectrum becomes
larger continuously as $\alpha$ decreases around
$\alpha\simeq1.550$, as shown in Fig.~\ref{fig:PointSpec}.
According to the analytical result in the neighborhood of a Hopf
bifurcation point (see pages~8-13 in Ref.~\cite{Springer.1984}), we may
expect that the amplitude of the limit-cycle solution gradually
increases as the real part of such eigenvalues increases.
Strictly speaking, the method in Ref.~\cite{Springer.1984} may not be
applied to the present spectral problem, because the method in
Ref.~\cite{Springer.1984} assumes that the linearized matrix of the
system is finite-dimensional but the present problem has a continuous
spectrum.
We again refer to this problem in Sec.~\ref{sec:Summary}.
Below, we will see that this increase of the amplitude causes the change
of the type~I breathing chimera to the type~II.

\begin{figure}[tb]
 \centering
 \includegraphics[width=86truemm]{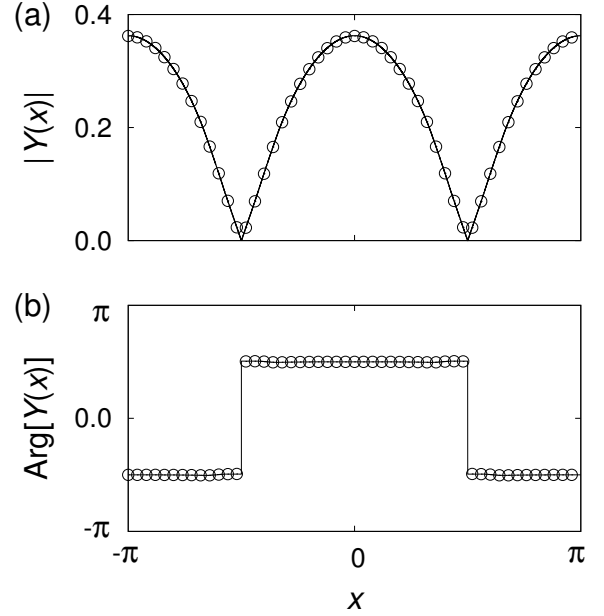}
 \caption{
 Local mean field of the type~I breathing chimera and the stationary
 chimera.
 Figures show (a)~the amplitude and (b)~the argument.
 Open circles denote $Y_0(x)$ of the type~I breathing chimera obtained
 by the numerical simulation of Eq.~(\ref{eq:DisPhaseOsc}) with
 $N=100000$, $\alpha=1.551$, and $r=0.620$.
 Those circles are plotted once every $2000$ oscillators.
 The solid line denotes the numerical solution $Y_{\rm st}(x)$ to the
 self-consistency equation~(\ref{eq:SelfCons}) at the same parameters.
 This solution corresponds to the unstable stationary chimera.
 }
 \label{fig:SelfCons-Stat}
\end{figure}

As mentioned in Sec.~\ref{sec:NumSim}, we have found the Hopf
bifurcation points at $r\simeq0.400$ and $r\simeq0.580$ between the
stationary chimera and the type~I breathing chimera by the linear
stability analysis.
Then, the absolute values of the imaginary parts of the point spectrum
are nearly equal to the breathing frequency $\delta$
\cite{PhysRevE.97.042212}.
This agrees with the occurrence of a supercritical Hopf bifurcation.
Therefore, for the type~I breathing chimeras with a small breathing
amplitude immediately after a Hopf bifurcation,
we can assume that the local order parameter $z(x,t)$ given by
Eq.~(\ref{eq:Breath-LO}) satisfies
\begin{equation}
 z_k(x) = O(\epsilon^{|k|}),
  \label{eq:First_LO}
\end{equation}
where $\epsilon$ is a small bifurcation parameter \cite{Springer.1984}.
Then, the local mean field $Y(x,t)$ also satisfies
\begin{equation}
 Y_k(x) = O(\epsilon^{|k|}),
  \label{eq:First_MF}
\end{equation}
from Eq.~(\ref{eq:MF_LO}).
For $k=0$, substituting Eqs.~(\ref{eq:First_LO}) and (\ref{eq:First_MF})
into Eq.~(\ref{eq:Sol_Breath-z}) and eliminating the $O(\epsilon^1)$
terms, Eq.~(\ref{eq:Sol_Breath-z}) is equivalent to the stationary case
Eqs.~(\ref{eq:Sol_Stat-z}) and (\ref{eq:EssSpec}), where
$A_0=e^{i\alpha}Y_0^*(x)$, $B_0=i\Delta_0$, and $C_0=e^{-i\alpha}Y_0(x)$.
Therefore, we have
\begin{equation}
 z_0(x) \simeq z_{\rm st}(x), \qquad Y_0(x) \simeq Y_{\rm st}(x),
  \label{eq:Breath-z_k0}
\end{equation}
where $z_{\rm st}(x)$ and $Y_{\rm st}(x)$ denote the quantities for the
unstable stationary chimera at the same parameters that remains after the
Hopf bifurcation of the stable stationary chimera.
These agree with the numerical result as shown in
Fig.~\ref{fig:SelfCons-Stat}.
$Y_0(x)$ of the type~I breathing chimera obtained by the numerical
simulation of Eq.~(\ref{eq:DisPhaseOsc}) and the numerical solution
$Y_{\rm st}(x)$ to the self-consistency equation~(\ref{eq:SelfCons})
look identical.

On the rotating frame with the frequency $\Omega$, $z(x,t)$ oscillates
around the center $z_0(x)$, and $z_k(x)\,e^{ik\delta t}$ for $k=\pm1$
are the main terms of oscillation for the type~I breathing chimera.
Substituting Eqs.~(\ref{eq:First_LO}) and (\ref{eq:First_MF}) into
Eq.~(\ref{eq:Sol_Breath-z}) for $k=\pm1$ and eliminating the
$O(\epsilon^2)$ terms, we obtain
\begin{equation}
 z_{\pm1}(x) \simeq
  \frac{-e^{-i\alpha}Y_{\pm1}(x) + e^{i\alpha}{z_0}^2(x)Y_{\mp1}^*(x)}
  {2[i\Delta_{\pm1} - e^{i\alpha}z_0(x)Y_0^*(x)]}.
  \label{eq:Breath-z_k1}
\end{equation}
$z_{\pm1}(x)$ are in the order of $\epsilon^1$ for almost all $x$, but in
the vicinity of $x_s$ they become larger than $O(\epsilon)$ and therefore
do not satisfy Eq.~(\ref{eq:Breath-z_k1}), if there exist specific points
$x=x_s$ satisfying
\begin{equation}
 i(\Omega + k\delta) = i\omega - e^{i\alpha} z_0(x) Y_0^*(x),
  \label{eq:Diverge}
\end{equation}
for $k=\pm1$, since the denominator of the right-hand side in
Eq.~(\ref{eq:Breath-z_k1}) becomes zero.
From Eq.~(\ref{eq:Breath-z_k0}), the right-hand side of
Eq.~(\ref{eq:Diverge}) agrees with Eq.~(\ref{eq:LO-conjMF}) for the
unstable stationary chimera in the order of $\epsilon^0$.
In incoherent regions, Eq.~(\ref{eq:LO-conjMF}) is purely imaginary and
its imaginary part corresponds to the average frequency, as mentioned in
Sec.~\ref{sec:Theory}.
Let us consider the case of $\Delta>0$.
For stationary chimeras, the average frequency of the coherent region is
equal to $\Omega$, which is the minimum value of the average frequency,
from Eq.~(\ref{eq:Stat-AveFreq}).
Since $\delta>0$, if $\Omega+\delta$ is within the range between the
maximum and the minimum of the average frequency, some points $x_s$
satisfying Eq.~(\ref{eq:Diverge}) for $k=1$ should exist.
On the other hand, if $\Delta<0$, $\Omega$ is the maximum value of the
average frequency.
Then, some $x_s$ satisfying Eq.~(\ref{eq:Diverge}) for $k=1$ exist under
the same condition of $\Omega+\delta$ since $\delta<0$.
Therefore, from Eq.~(\ref{eq:Stat-AveFreq}), if the breathing frequency
$\delta$ satisfies the condition
\begin{equation}
 0 < |\delta|
  \leq {\rm max} \{ |\Delta \sqrt{1 - (|Y_{\rm st}(x)|/\Delta)^2}| \},
  \label{eq:xs-Cond}
\end{equation}
in incoherent regions ($\Delta>|Y_{\rm st}(x)|$), some specific points
$x_s$ exist, and $|z_1(x_s)|$ becomes larger sharply than other points
$x$.
We note that $|z_1(x_s)|$ does not diverge to infinity.
The function $z_{\pm 1}(x)$ is determined by Eq.~(\ref{eq:Sol_Breath-z}),
but can be approximately found by Eq.~(\ref{eq:Breath-z_k1}) for all
$x\in[-\pi,\pi)$ such that the denominator in Eq.~(\ref{eq:Breath-z_k1})
is separated from zero.
At $x=x_s$, the denominator in Eq.~(\ref{eq:Breath-z_k1}) for $k=1$ is
zero.
Then, $z_{\pm1}(x_s)$ does not obey Eq.~(\ref{eq:Breath-z_k1}).
However, the correct values of $z_{\pm1}(x_s)$ still can be found from
Eq.~(\ref{eq:Sol_Breath-z}), and $|z_1(x_s)|$ practically becomes a
large finite value.
In our numerical simulations ($\omega=0$) presented here, we observed
$\Delta>0$ and therefore Eq.~(\ref{eq:xs-Cond}) becomes
\begin{equation}
 0 < \delta \leq \Delta \, ( = -\Omega ),
  \label{eq:xs-Cond_pre}
\end{equation}
since the minimum of $|Y_{\rm st}(x)|$ is zero as shown in
Fig.~\ref{fig:SelfCons-Stat}.

From the existence of specific points $x_s$, we can explain that the
type~I changes to the type~II by increasing the breathing amplitude, as
follows.
After the Hopf bifurcation from stationary chimeras, there appear the
type~I breathing chimeras with a small breathing amplitude.
This amplitude is mainly characterized by $z_{\pm1}(x)$, which are very
small for almost $x$.
However, $z_1(x)$ is large only at $x_s$.
As increasing the breathing amplitude by leaving the bifurcation point,
$z_{\pm1}(x)$ gradually becomes large.
By the increase in $z_{\pm1}(x)$, especially $z_1(x_s)$, $z(x,t)$
reaches the upper limit $|z(x,t)|=1$ at $x_s$, e.g., for
$\alpha\simeq1.550$ and $r=0.620$.
When $\alpha$ decreases further from $\alpha=1.550$ with fixed
$r=0.620$, $z_1(x_s)$ cannot become large anymore.
Instead, the second coherent regions with average frequency
$\Omega+\delta$ emerge around $x_s$ with increasing the amplitude; in
other words, the type~II breathing chimera appears.

\begin{figure}[tb]
 \centering
 \includegraphics[width=86truemm]{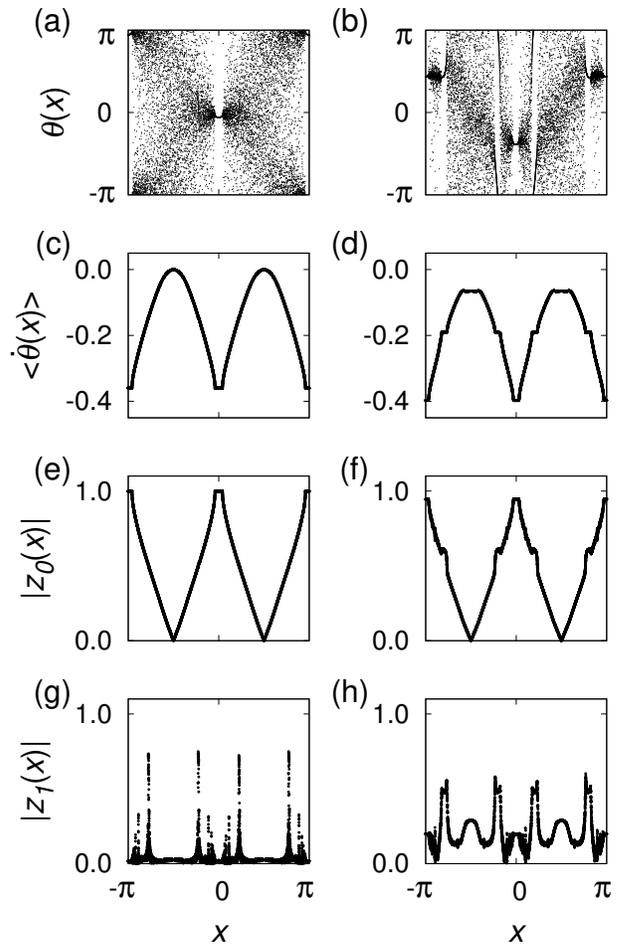}
 \caption{
 Comparison between the two types of breathing chimeras for the
 numerical simulation of Eq.~(\ref{eq:DisPhaseOsc}) with $N=100000$ and
 $r=0.620$.
 The left column denotes the type~I for $\alpha=1.551$, and the right
 column denotes the type~II for $\alpha=1.549$.
 Figures~(a) and (b) show the snapshot of the phase.
 Figures~(c) and (d) show the profile of the average frequency.
 Figures~(e) and (f) and figures~(g) and (h) show the amplitudes of
 $z_0(x)$ and $z_1(x)$, respectively.
 All the figures are plotted once every $10$ oscillators.
 Note that the appearance of the types~I and II is switched at
 $\alpha\simeq1.550$.
 }
 \label{fig:AVE-LO}
\end{figure}

Let us confirm this scenario by numerical simulations of
Eq.~(\ref{eq:DisPhaseOsc}).
Figure~\ref{fig:AVE-LO} shows comparison between the two types of
breathing chimeras near the bifurcation between them.
For the type~I breathing chimera for $r=0.620$ and $\alpha=1.551$ (see
the left column in Fig.~\ref{fig:AVE-LO}), we obtained
$\Omega\simeq-0.3602$ and $\delta\simeq0.2151$, then from
Eq.~(\ref{eq:Diverge}) we can see that
$\langle\dot{\theta}(x_s,t)\rangle=\Omega+\delta$ is established for
$k=1$ at, e.g., $x_s\simeq0.705$.
Such a profile as $|z_1(x)|$ nearly diverges can often be seen just
before the bifurcation to the type~II.
As shown in Fig.~\ref{fig:AVE-LO}(g), $|z_1(x)|$ is very small for
almost all $x$ but nearly diverges at the points $x_s$.

For the type~II breathing chimeras, the local order parameter does not
satisfy Eq.~(\ref{eq:First_LO}), because $z_0(x)$ as shown in
Fig.~\ref{fig:AVE-LO}(f) clearly differs from $z_{\rm st}(x)$
[$\simeq z_0(x)$ for the type~I]; that is, Eq.~(\ref{eq:Breath-z_k0}) is
not satisfied.
Therefore, it turns out that the breathing amplitude for type~II is
larger than that for type~I.
When Figs.~\ref{fig:AVE-LO}(a) and \ref{fig:AVE-LO}(b) are compared, we
find that a part of the incoherent region suddenly changes to the second
coherent region.
Then, it is observed that the second coherent regions for type~II emerge
at the same points as $x_s$ for type~I and have the average frequency
$\Omega+\delta$ obtained from the simulation results
$\Omega\simeq-0.3974$ and $\delta\simeq0.2067$.
From this result, our scenario is shown to be valid.
As shown in Fig.~\ref{fig:StabRegion}, we do not observe that the type~I
breathing chimeras for $r<0.400$ change to the type~II.
This seems to be because the amplitude increase is smaller than that for
$r>0.600$.

For $|k|\geq2$, we can obtain $z_k(x)$ of the order of $\epsilon^{|k|}$
similar to Eq.~(\ref{eq:Breath-z_k1}) and the same condition as
Eq.~(\ref{eq:Diverge}).
Therefore, there can also exist special points $x_s$ satisfying
Eq.~(\ref{eq:Diverge}), if $\Omega+k\delta$ is within the range between
the maximum and minimum of the average frequency.
In other words, if the breathing frequency $\delta$ satisfies the
condition
\begin{equation}
 0 < k|\delta|
  \leq {\rm max} \{ |\Delta \sqrt{1 - (|Y_{\rm st}(x)|/\Delta)^2}| \},
  \label{eq:xs-Cond-k}
\end{equation}
the type~II breathing chimera has $(k+1)$-th coherent regions with the
average frequency $\Omega+k\delta$.
In the present case, $\delta$ does not satisfies
Eq.~(\ref{eq:xs-Cond-k}) except for $k=1$, so the type~II breathing
chimeras cannot have the third and greater coherent regions.
However, the type~II breathing chimeras in Refs.~\cite{JetpLett.106.393,
Chaos.28.045101} appear to have the second and third coherent regions,
though the system used in Refs.~\cite{JetpLett.106.393, Chaos.28.045101}
includes phase lag parameter heterogeneity.
We emphasize that our analytical theory and scenario can be applied to
the system with phase lag parameter heterogeneity only by replacing
$\alpha$.
As mentioned above, $|\delta|$ is nearly equal to the absolute value
of the imaginary parts of the point spectrum in the neighborhood of a
Hopf bifurcation point.
Therefore, when the type~I breathing chimera is bifurcated via Hopf
bifurcation, it is already determined whether the type~II breathing
chimera has the second or greater coherent regions.

\section{Self-Consistent Analysis}
\label{sec:SelfCons}

Finally, we propose a self-consistent analysis for breathing chimeras.
As mentioned in Sec.~\ref{sec:Theory},
Eqs.~(\ref{eq:Sol_Breath-z})-(\ref{eq:Quad-Eq_C}) are the
self-consistency equations of $\{z_k(x)\}$.
In this section, we numerically solve them, especially, for the type~II
breathing chimera.

Equations.~(\ref{eq:Sol_Breath-z})-(\ref{eq:Quad-Eq_C}) are composed of
one complex equation for every $k$.
Therefore, we need two additional conditions to obtain the solution
because there are unknown complex functions $\{z_k(x)\}$ and two real
unknowns $\Omega$ and $\delta$ to be determined.
Unlike the breathing chimeras, the self-consistency
equation~(\ref{eq:SelfCons}) for stationary chimeras has one unknown
complex function $Y_{\rm st}(x)$ and one real unknown $\Omega$, and an
additional condition obtained from the invariance of
Eq.~(\ref{eq:SelfCons}) under any rotation of the argument of
$Y_{\rm st}(x)$ leads to solving the self-consistency
equation\cite{PhysRevLett.93.174102, IJBC.16.21, PhysRevLett.100.144102,
PhysRevE.90.022919, PhysRevE.97.042212}; for example,
${\rm Arg}[Y_{\rm st}(0)]=0$ is chosen.

Equation.~(\ref{eq:LO-conjMF}) can be utilized for obtaining the
additional real conditions to determine $\Omega$ and $\delta$.
The average frequencies of the first and second coherent regions are
equal to $\Omega$ and $\Omega+\delta$, respectively.
Moreover, the coherent region satisfies the stability condition
${\rm Re}f(x)<0$, and ${\rm Re}f(x)$ has a minimal value in every
coherent region, as shown in Fig.~\ref{fig:LO-conjMF}.
Let $x_{c1}$ and $x_{c2}$ be the minimal points of ${\rm Re}f(x)$
corresponding to the first and second coherent regions, respectively.
Then, the frequencies $\Omega$ and $\delta$ are given by
\begin{equation}
 \Omega = {\rm Im}f(x_{c1}),
  \label{eq:Omega}
\end{equation}
\begin{equation}
 \delta = {\rm Im}f(x_{c2}) - {\rm Im}f(x_{c1}).
  \label{eq:delta}
\end{equation}
Note that Eq.~(\ref{eq:Omega}) is also established for stationary
chimeras.
In the following, we regard
Eqs.~(\ref{eq:Sol_Breath-z})-(\ref{eq:Quad-Eq_C}) and
Eqs.~(\ref{eq:Omega}) and (\ref{eq:delta}) as the complete
self-consistency equations for the type~II breathing chimeras.

There are a few important points to solve the self-consistency
equations numerically.
First, we truncate $\{z_k(x)\}$ to $k\in[-10,10]$, assuming that
$z_k(x)$ for sufficiently large $|k|$ is small enough not to affect the
other $z_k(x)$.
That is confirmed from the results of the numerical simulation of
Eq.~(\ref{eq:DisPhaseOsc}).

The second point is the selection method of the argument of the square
root in Eq.~(\ref{eq:Sol_Breath-z}).
Equation.~(\ref{eq:Sol_Breath-z}) can produce two solutions according to
this selection.
In our numerical computation of the two solutions for all $k$, we have
found that the orders of these two solutions are greatly different
except for the first coherent regions for $k=0$ and the second coherent
regions for $k=1$.
In that case, the larger one is easily rejected because of the condition
$|z(x,t)|<1$.
The problem is the exceptional case where the orders of the two
solutions are not so different.
Then, one of the two solutions corresponds to the stable solution and
the other does not.
That can be shown as follows.
Because Eqs.~(\ref{eq:Sol_Breath-z})-(\ref{eq:Quad-Eq_C}) are
transformed to
\begin{eqnarray}
 \{ {B_k}^2(x)-A_k(x)C_k(x) \}^{\frac{1}{2}}
 &=& -i\Delta_k + e^{i\alpha}
  \sum_{k=-\infty}^{\infty} z_k(x) \, Y_k^*(x), \nonumber \\
 &=& i( \Omega + k\delta ) - f(x),
  \label{eq:SquareRoot}
\end{eqnarray}
where $f(x)$ is the same function in Eq.~(\ref{eq:LO-conjMF}), we have
\begin{equation}
 f(x) = i( \Omega + k\delta )
  -\{ {B_k}^2(x)-A_k(x)C_k(x) \}^{\frac{1}{2}}.
  \label{eq:SquareRoot-f}
\end{equation}
It is interesting that the right-hand side of
Eq.~(\ref{eq:SquareRoot-f}) should be independent of $k$.
Therefore, since ${\rm Im}f(x)=\Omega$ in the first coherent regions,
the square root becomes the real number for $k=0$, and either one
corresponding to ${\rm Re}f(x)<0$ should be selected from the stability
in the coherent regions.
The case in the second coherent regions for $k=1$ is the same as above.

In this way, we can select the stable solution to
Eq.~(\ref{eq:Sol_Breath-z}) at almost all $x$ for $k=0,1$.
However, the stable and unstable solutions are too close to be
distinguished around the boundaries between the coherent and incoherent
regions since ${\rm Re}f(x)\simeq0$.
To solve this problem, we use the following method.
For example, let us consider the boundaries between the first coherent
and incoherent regions for $k=0$.
Substituting Eq.~(\ref{eq:SquareRoot}) into
Eqs.~(\ref{eq:Sol_Breath-z})-(\ref{eq:Quad-Eq_C}), we have
\begin{equation}
 z_k(x) = [ B_k(x) + i( \Omega + k\delta ) - f(x) ] / A_k(x).
  \label{eq:SC-Breath_f}
\end{equation}
This equation is also derived from Eq.~(\ref{eq:LO-conjMF}) directly.
Because the branch of the square root for $k\neq0$ is easily selected by
the orders of the two solutions, the right-hand side of
Eq.~(\ref{eq:SquareRoot-f}) can be computed for a specific $k\neq0$.
When it is difficult to distinguish the two solutions for $k=0$, we may
use Eq.~(\ref{eq:SquareRoot-f}) for $k\neq0$ as $f(x)$ in
Eq.~(\ref{eq:SC-Breath_f}).

\begin{figure}[tb]
 \centering
 \includegraphics[width=86truemm]{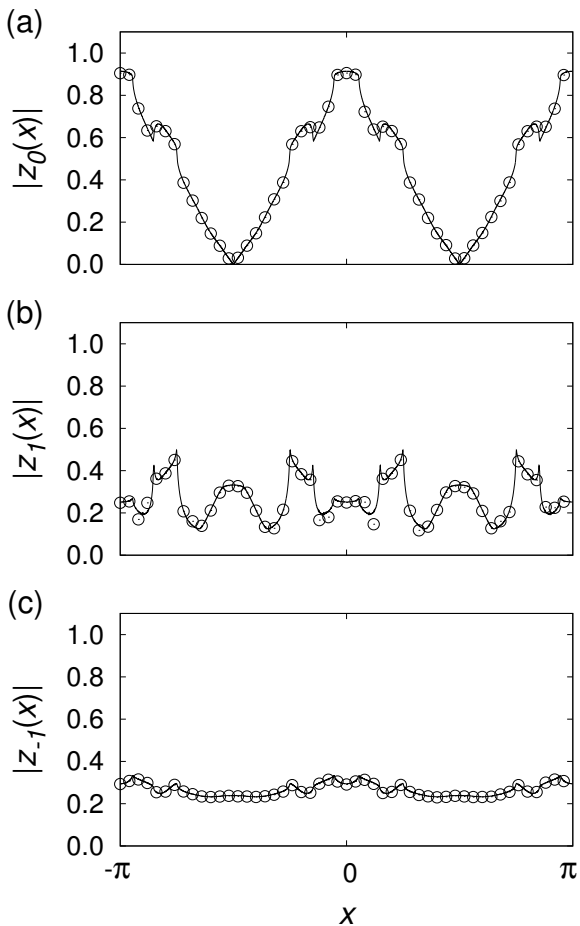}
 \caption{
 Local order parameter of the type~II breathing chimera for
 $\alpha=1.500$ and $r=0.600$ corresponding to Fig.~\ref{fig:Type2}.
 Figures show the amplitude of (a)~$z_0(x)$, (b)~$z_1(x)$, and
 (c)~$z_{-1}(x)$.
 Open circles denote the values obtained by the numerical simulation of
 Eq.~(\ref{eq:DisPhaseOsc}) with $N=100000$, and are plotted once every
 $2000$ oscillators.
 The solid line denotes the numerical solution to the self-consistency
 equations~(\ref{eq:Sol_Breath-z})-(\ref{eq:Quad-Eq_C}),
 (\ref{eq:Omega}), and (\ref{eq:delta}).
 }
 \label{fig:SelfCons-Breath}
\end{figure}

Figure~\ref{fig:SelfCons-Breath} shows numerical solutions to the
self-consistency equations~(\ref{eq:Sol_Breath-z})-(\ref{eq:Quad-Eq_C}),
(\ref{eq:Omega}), and (\ref{eq:delta}).
At first, we tried to numerically solve the self-consistency equations
by the simple iteration method, where unknown variables $\{z_k(x)\}$,
$\Omega$, and $\delta$ are substituted into the right-hand side of the
equations and are regenerated from the left-hand side.
However, we could not obtain a solution of the type~II breathing
chimeras because the variables have not converged even by using various
initial conditions.
Instead, we have applied Steffensen's method
\cite{doi:10.1080/03461238.1933.10419209} to the regeneration of every
variable and have succeeded in obtaining the correct numerical solution.
Open circles in Fig.~\ref{fig:SelfCons-Breath} denote $z_k(x)$ obtained
by the numerical simulation of Eq.~(\ref{eq:DisPhaseOsc}).
We used them as the initial condition for solving the self-consistency
equations.
The results from the numerical simulation and the self-consistency
equations look like almost identical.
Although it may seem that they are not identical in a part of $|z_1(x)|$,
that is caused by the numerical error due to the finite-size effects of
the simulations.
We expect that more extensive simulations improve this discrepancy.
We succeeded in obtaining the solution to the self-consistency equations
by using an initial condition that is very close to the correct
solution.
However, when other initial conditions were used, the correct solution
could not be obtained since the variables have not converged.
This may be a weak point of our numerical method.

\section{Summary}
\label{sec:Summary}

We have studied breathing chimera states in one-dimensional nonlocally
coupled phase oscillators.
First, we have found breathing chimeras in numerical simulations.
The breathing chimeras are characterized by the temporally oscillating
global order parameter and classified into two types by observing the
average frequencies of the coherent regions.
While type~I contains the coherent regions with a common average
frequency similarly to the stationary chimera, type~II contains the
coherent regions with different average frequencies.
Type~II breathing chimeras are also obtained for
Eq.~(\ref{eq:PhaseOsc}) with phase lag parameter heterogeneity
\cite{JetpLett.106.393, Chaos.28.045101}.

Next, we have assumed that the local order parameter $z(x,t)$ takes the
form of Eq.~(\ref{eq:Breath-LO}) instead of Eq.~(\ref{eq:Stat-LO}) as in
many previous works, and analytically discussed breathing chimeras.
Moreover, we have derived the self-consistency
equations~(\ref{eq:Sol_Breath-z})-(\ref{eq:Quad-Eq_C}) and the important
complex function Eq.~(\ref{eq:LO-conjMF}), whose imaginary and real
parts denote the average frequency and the local linear stability,
respectively.
They turns out to be very useful to analyze breathing chimeras.

We have shown that the type~I breathing chimera changes to type~II by
increasing the breathing amplitude.
This means that the type~I breathing chimera looks the same as the
stationary chimera since the breathing amplitude is small but the second
coherent regions emerge in the incoherent regions as that amplitude
becomes larger.
Such a bifurcation, that new coherent regions emerge in the incoherent
regions, has been reported in a few systems that are different from phase
oscillators \cite{PhysRevLett.110.224101, PLoSONE.12.e0187067}.
However, the mechanism of that bifurcation in the other systems is
unclear.

In the present paper, we applied the analytical result in
Ref.~\cite{Springer.1984} to the spectral problem corresponding to the
linearization of Eq.~(\ref{eq:z-Eq}).
However, this method may not rigorously be justified in $N\to\infty$.
To understand breathing chimera states more precisely, we need to
analyze them by a more sophisticated perturbation theory for
infinite-dimensional systems \cite{Springer.1995}.

Finally, we have numerically solved the self-consistency
equations~(\ref{eq:Sol_Breath-z})-(\ref{eq:Quad-Eq_C}).
Then, the frequencies $\Omega$ and $\delta$ are formulated as
Eqs.~(\ref{eq:Omega}) and (\ref{eq:delta}), respectively.
Our numerical method has succeeded in solving them, but it is necessary
to use the initial condition that is very close to the correct solution.
To obtain the solution more easily, we need to improve the present
method in future.

\bibliography{Reference}

\end{document}